\documentclass[a4paper]{llncs}

\usepackage[T1]{fontenc}
\usepackage{lmodern}
\usepackage[scaled=.8]{beramono}
\usepackage{latexsym}
\usepackage{textcomp}

\usepackage{todonotes}

\usepackage[utf8]{inputenc}
\usepackage{hyperref}
\usepackage[english]{babel} 
\usepackage{mathtools}
\usepackage{amssymb}
\usepackage{listings}
\usepackage{amsmath}
\usepackage{amsfonts}
\usepackage{mathabx}

\usepackage{url}
\newcommand{\keywords}[1]{\par\addvspace\baselineskip
\noindent\keywordname\enspace\ignorespaces#1}

\usepackage{graphicx}
\graphicspath{ {images/} }
\usepackage{enumerate}
\usepackage{caption}
\usepackage{subcaption}

\usepackage{enumitem}

 \title{Interest-based RDF Update Propagation}
 \author{Kemele M. Endris, Sidra Faisal, Fabrizio Orlandi, S\"{o}ren Auer, Simon Scerri}
\institute{University of Bonn \& Fraunhofer IAIS, Bonn, Germany\\
\email{\{lastname\}@cs.uni-bonn.de} }

\begin{document}
\maketitle
\vspace{-1em}
\begin{abstract}

Many LOD datasets, such as DBpedia and LinkedGeoData, are voluminous and process large amounts of requests from diverse applications. 
Many data products and services rely on full or partial local LOD replications to ensure faster querying and processing. 
While such replicas enhance the flexibility of information sharing and integration infrastructures, they also introduce data duplication with all the associated undesirable consequences. 
Given the evolving nature of the original and authoritative datasets, to ensure consistent and up-to-date replicas frequent replacements are required at a great cost. In this paper, we introduce an approach for interest-based RDF update propagation, which propagates only interesting parts of updates from the source to the target dataset. 
Effectively, this enables remote applications to `subscribe' to relevant datasets and consistently reflect the necessary changes locally without the need to frequently replace the entire dataset (or a relevant subset). 
Our approach is based on a formal definition for graph-pattern-based interest expressions that is used to filter interesting parts of updates from the source. 
We implement the approach in the iRap framework and perform a comprehensive evaluation based on DBpedia Live updates, to confirm the validity and value of our approach.

\end{abstract}
\vspace{-2.5em}
\vspace{-1em}
\section{Introduction}
\vspace{-1em}
In recent years, there has been an increasing number of structured data published on the Web as a Linked Open Data (LOD).
Last years assessment of the size of the LOD cloud\footnote{\url{http://linkeddatacatalog.dws.informatik.uni-mannheim.de/state/}} for example reported more than 1.000 published datasets comprising almost 100 Billion triples.
Methods for accessing LOD are SPARQL endpoints, Linked Data resource documents or data dumps.
Many of these datasets, such as DBpedia and LinkedGeoData, are voluminous and process large amount of requests from diverse applications.
Providing services on top of these datasets is becoming a challenge due to the lack of service levels regarding the availability of datasets and restrictions imposed by the publisher on the type of query forms and number of results.  

Replication of Linked Data datasets enhances flexibility of information sharing and integration infrastructures. 
Since hosting a replica of large datasets, such as DBpedia and LinkedGeoData, is costly, organizations might want to host only a relevant subset of the data, for example, using approaches such as \emph{RDFSlice}~\cite{MarxSAN13}.
However, due to the evolving nature of these datasets in terms of content and ontology, maintaining a consistent and up-to-date replica of the relevant data is a major challenge.
Resources in a dataset might be added, updated, or removed. 
The frequency of such changes depends on the type of data stored in a dataset. 
For example, sensor data or geolocation data from mobile devices changes more frequently than archival data. 
These changes should be dealt with by Linked Data consumption applications in order to keep local repositories consistent.

\begin{figure}[tb]
\vspace{-2em}
	\centering	
	\includegraphics[width=1\textwidth]{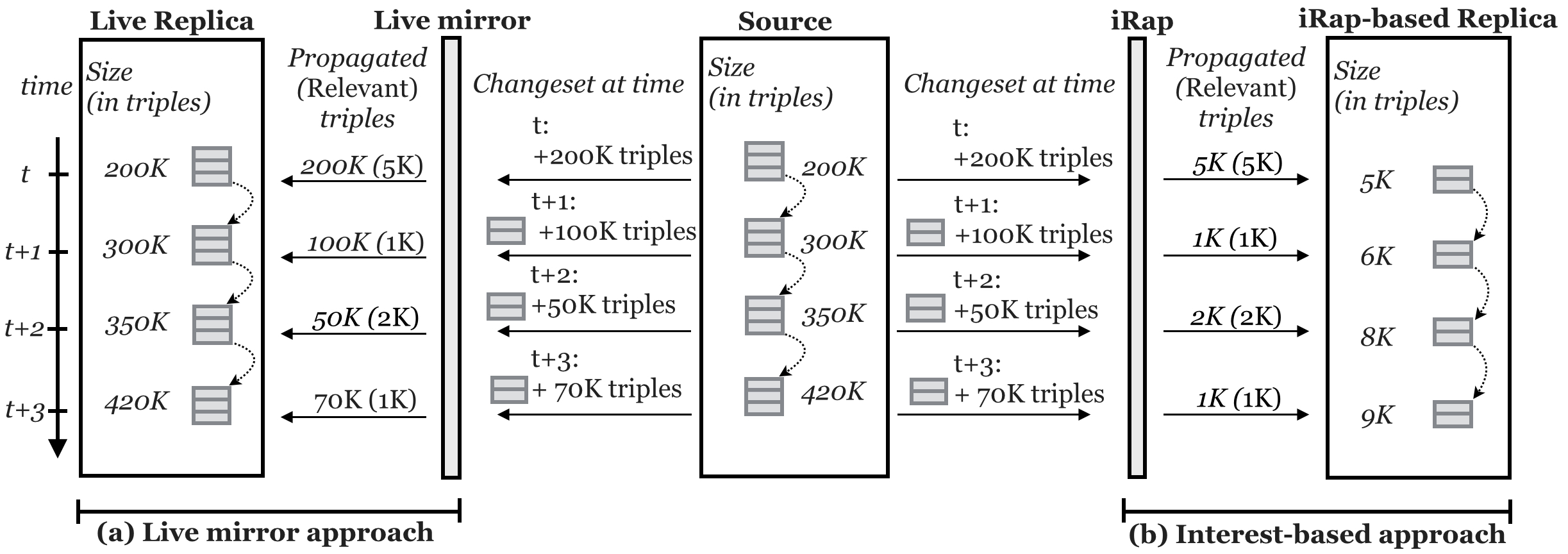}	
	\caption{Changeset propagation approaches: right part -- Interest-based replica (iRap Replica); left part -- Live mirror replica (Live Replica)}
	\label{fig:propapproaches}
	\vspace{-1.7em}
\end{figure}

Typically, a dataset mirror application propagates a changeset, published by the source dataset, to a target dataset. 
For example, the \emph{DBpedia Live mirror tool}\footnote{\url{https://github.com/dbpedia/dbpedia-live-mirror}} propagates all changesets to a target dataset, so that at time $t$ the target dataset contains the same triples as the source dataset. 
However, for example, an application interested in athletes uses only 268,773 out of 364,810,370 instances of the English DBpedia 2014 dataset. 
An interest-based update propagation could significantly reduce the amount of data to be shipped and managed at the application side and thus lower the barrier for the deployment of Linked Data applications.
In this paper, we present an approach for interest-based update propagation, which is based on the specification of data interests by a target application.
Based on such interest expressions all updates are evaluates at the source and only those are shipped to the target application, which are either directly interesting or could become interesting in subsequent updates.
We provide a thorough formalization of our approach.
~\autoref{fig:propapproaches} shows that propagation of unfiltered data from Source to Target-2 (in part b) syncing the complete changeset irrespective of the relevant or useful data whereas, the propagation of filtered data using iRap from Source to Target-1 (for part a) transfers only relevant data. 
Our evaluation shows, that the data required to be transfered and handled by applications can be reduced by several orders of magnitude thus substantially lowering the re-usage barrier for Linked Data.

The article is structured as follows:
\autoref{sec:formalization} extensively describes the formalization for our framework.
\autoref{sec:iRapImpl} and \autoref{sec:evaluation} discusses the implementation and evaluation of the iRap framework in detail.
\autoref{sec:relatedwork} describes the related work. 
Finally,~\autoref{sec:conclusion} concludes and proposes directions for future work.

\vspace{-1em}
\section{Formalization of Interest-based RDF Updates}\label{sec:formalization}

\begin{figure}[b!]
	\centering	
	\vspace{-1em}
	\includegraphics[width=.7\textwidth]{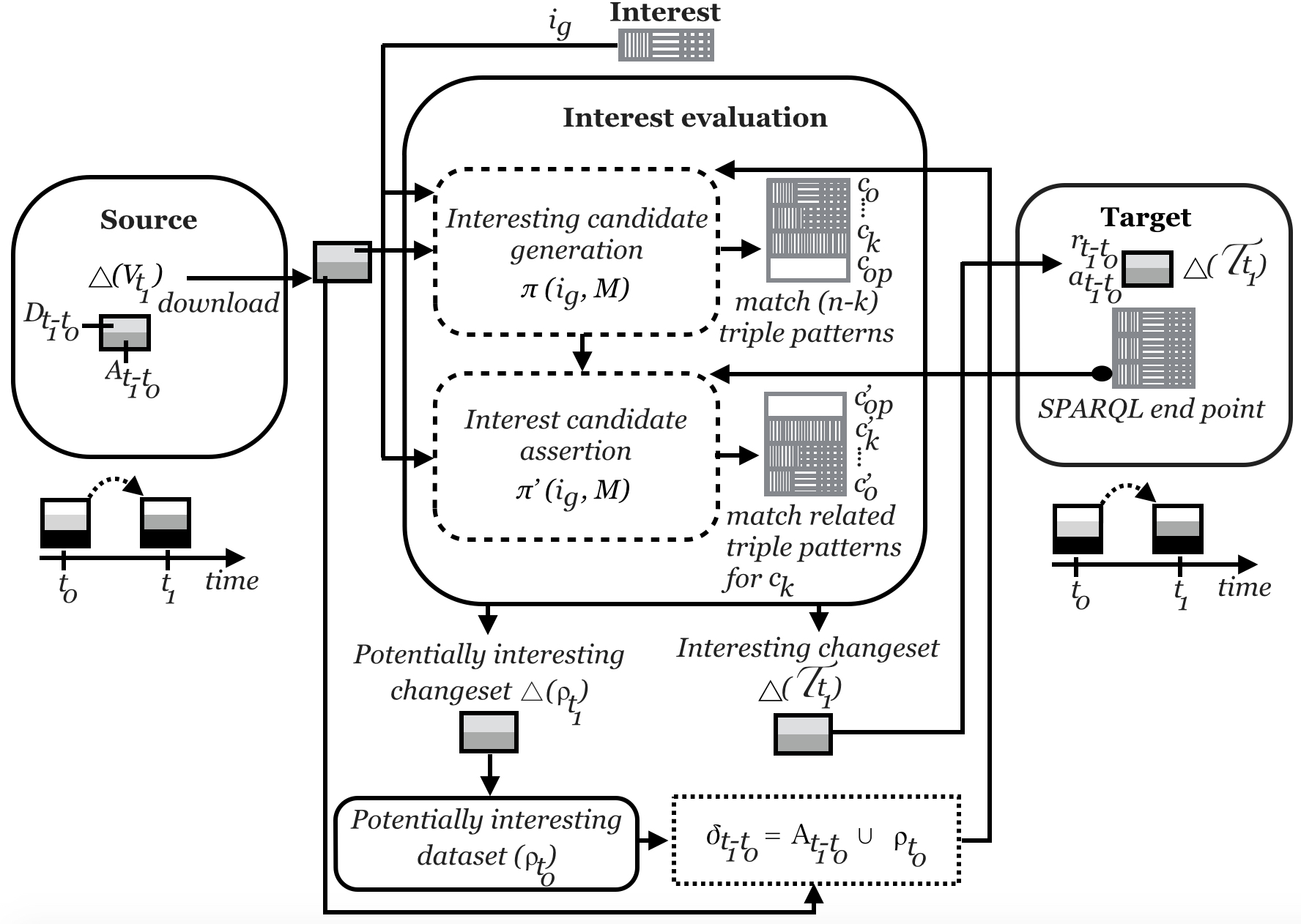}
	\caption{Formalization overview of the interest-based RDF update propagation.}
	\label{fig:approach}
	\vspace{-0.5em}
\end{figure}

\autoref{fig:approach} illustrates the overall interest-based RDF Update Propagation approach; summarizing the concepts defined through the formalization. 
Interest evaluation takes place over the input set of deleted ($D_{t_1-t_0}$) and added ($A{t_1-t_0}$) triples from the source dataset ($V_{t_1}$) in between time interval $(t_0,t_1)$.
Since updates can not only contain interesting and uninteresting parts but also triples, which can become potentially interesting along with subsequent updates, we have to compute and store these sets of potentially interesting triples and take them in subsequent update assessments into account.

For our formalization we will use the standard notations \textbf{I}, \textbf{B}, \textbf{L} and \textit{Var} for the disjoint sets of all IRIs, blank nodes, literals (typed and untyped) and variables respectively. 
An \textit{RDF graph} V is a finite set of RDF triples, i.e, $V\subset^F$ (\textbf{I}$\cup$\textbf{B}) x \textbf{I} x (\textbf{I}$\cup$\textbf{B}$\cup$\textbf{L}). 
In this paper we use the terms RDF graph, \textit{RDF dataset}, and \textit{dataset} interchangeably.

\begin{definition}[Evolving Dataset]\label{def:evolvingDataset}
An evolving dataset $V^g$ is a dataset identified using the persistent IRI $g$ whose content changes over time. 
$V^g_t$ denotes a specific revision of $V^g$ at a particular time t.
For simplicity, we will just refer to $V_t$ instead of $V^g_t$.
\end{definition}
\vspace{-1.5em}
\begin{definition}[BGP]\label{def:BGP}
A SPARQL basic graph pattern (BGP) expression is defined recursively as follows:
\begin{enumerate}[nosep]
\item a triple pattern tp $\in (\textbf{I} \cup \textbf{B} \cup \textit{Var}) \;x\; (\textbf{I} \cup \textit{Var})\;x\; (\textbf{I} \cup \textbf{B} \cup \textbf{L} \cup \textit{Var})$ is a BGP
\item the expression (P1 AND P2) is a BGP, where P1 and P2 are themselves BGPs
\item the expression (P FILTER E) is a BGP, where P is a BGP and E is a SPARQL filter expression that evaluates to boolean value.
\end{enumerate}
\end{definition}
\vspace{-1.5em}
\begin{definition}[Non-disjoint BGP] \label{def:nondisjointBGP}
A non-disjoint BGP is a BGP that represents a connected graph.
\end{definition}
\vspace{-0.5em}
 An optional graph pattern (OGP) is syntactically specified with the OPTIONAL keyword applied to a graph pattern. A set of triple patterns in a BGP must match for there to be a solution whereas triple patterns in OGP may extend the solution but their non-binding nature means that they cannot reject it.~\cite{SparqlQuery}

\begin{definition}[Partial Matches] \label{def:partialMatching}
Partial matches are a set of triples that does not fully match the BGP but matches at least one triple pattern in BGP or OGP of a query.
\end{definition}

Triples added to, and removed from, an evolving dataset within a time-frame are called \textit{changeset} for a dataset within that time-frame. 
\begin{definition}[Changeset]\label{def:Changeset}
Let $V_{t_1}$ be an evolving dataset at time $t_1$. 
A changeset $\Delta(V_{t_1})$, between $V_{t_0}$ and $V_{t_1}$, where $t_0 < t_1$, is defined as:
\vspace{-0.3em}
\[\Delta(V_{t_1})= \left\langle D_{t_1-t_0}, A_{t_1-t_0}\right\rangle\] where:
$D_{t_1-t_0}$ is a set of removed triples from $V_{t_0}$ between time-points $t_0$ and $t_1$, and $A_{t_1-t_0}$ is a set of added triples to $V_{t_0}$ between time-points $t_0$ and $t_1$.
\end{definition}
Changesets can be computed using the difference between two versions of the RDF dataset.
The result of this computation gives the removed triples, $D_{t_1-t_0} = V_0 \setminus V_1$, and added triples, $A_{t_1-t_0}= V_1 \setminus V_0$, between given dataset revisions $V_{t_0}$ and $V_{t_1}$.
Datasets can be accompanied with a tool that publishes changesets at real-time, so that users can download these and synchronize their local replicas. 
For instance, DBpedia publishes updates in a public changesets folder\footnote{\url{http://live.dbpedia.org/changesets/}}. 
\begin{example} \label{exmp:changeset}
Let us assume the following two files\footnote{prefixes can be checked in \url{http://prefix.cc/}} are being published by the DBpedia Live extractor for the changes made on Feb 06, 2015 between 05:00 PM ($t_0$) and 05:02 PM ($t_1$):
\vspace{-0.5em}
\begin{figure}[tbh]
\begin{subfigure}[h]{0.46\textwidth}
\begin{lstlisting}[caption={File 000001.removed.nt}]
dbr:Marcel             dbp:goals     1 .
dbr:Marcel             dbo:team      dbr:FNFT .
dbr:Tim%02             foaf:name  
                         "Tim Berners-Lee" .
dbr:Cristiano_Ronaldo  dbo:goals    96 .
\end{lstlisting}
\end{subfigure}
\hspace{.2cm}
\begin{subfigure}[h]{0.5\textwidth}
\begin{lstlisting}[caption={File 000001.added.nt}]
dbr:Cristiano_Ronaldo  dbo:goals 216 . 
dbr:Barack_Obama       foaf:name "Barack Obama" .
dbr:Barack_Obama       foaf:homepage  
                      "http://www.barackobama.com/" .
dbr:Rio_Ferdinand      a         foaf:Person .
dbr:Rio_Ferdinand      a         dbo:Athlete .
dbr:Rio_Ferdinand      dbp:goals 2 .
dbr:Arvid_Smit         a         dbo:Athlete .
\end{lstlisting}
\end{subfigure}
\end{figure}

A changeset $\Delta(V_{t_1})$ for the DBpedia Live dataset between $t_0$ and $t_1$, contains $D_{05:02-05:00}=000001.removed.nt$ and  $A_{05:02-05:00}=000001.added.nt$. 
That is,  $\Delta(V_{05:02})=\left\langle 000001.removed.nt, 000001.added.nt\right\rangle$
\end{example}

\begin{definition}\label{def:ChangesetPropagation}
\textbf{(Changeset Propagation)} A changeset propagation is a function $\upsilon$ that transforms a given dataset $V_{t_0}$ to a new dataset  $V_{t_1}$ by applying a changeset, $\Delta(V_{t_1})$.
That is:
\vspace{-0.5em}
\[ \upsilon(V_{t_0}, \Delta(V_{t_1}) ) = \lgroup V_{t_0} \backslash D_{t_1-t_0}\rgroup \cup A_{t_1-t_0} = V_{t_1} \]
\end{definition}
\vspace{-0.3em}
The changeset propagation function $\upsilon$, for example, deletes the triples in 000001.removed.nt from the target dataset and then inserts all triples from 000001.added.nt. 
This order of operation (deleted first) ensures that inserted triples are not removed again immediately. 

If an organization maintaining a replica wants to host only a subset of the original dataset it needs to obtain only relevant updates for this subset.
For that purpose, we specify \textit{interests} to subscribe to `interesting' changes only.
During interest registration, an organization provides information about the source dataset to synchronize with, a target dataset endpoint that supports SPARQL Update\footnote{\url{http://www.w3.org/TR/sparql11-update/}} to propagate interesting changes, and an interest expression to select relevant parts of a changeset.
Below, we present a formal definition for \textit{interest expression} over an evolving dataset.
\begin{definition}[Interest Expression]\label{def:InterestExp}
An \textit{interest expression} over an evolving dataset, $V^g_t$, is defined as:
\vspace{-0.5em}
 \[i_g = \left\langle \tau, b, op\right\rangle\] where
$g$ is an IRI identifying an evolving RDF dataset $V^g$, 
$\tau$ is an IRI identifying the target dataset endpoint, 
$b$ is a non-disjoint BGP, and
$op$ is an optional graph pattern (OGP) connected to $b$.
\end{definition}

\begin{example} \label{exmp:interest}
An interest expression for a list of an athlete with information about goals scored, and optionally their homepage, is expressed as follows: 
\begin{itemize}[nosep]
	\item g = \textit{http://live.dbpedia.org/changesets}
	\item $\tau$ = \textit{http://localhost:3030/target/sparql}
	\item b = \texttt{\{ ?a	a  dbo:Athlete .  ?a	dbp:goals  ?goals . \}}
	\item op = \texttt{\{ ?a foaf:homepage  ?page . \}}
\end{itemize}
The equivalent interest expression SPARQL query will be:
\begin{lstlisting}[language=sparql,escapechar=@,morekeywords={STRSTARTS}, frame=none]
SELECT * WHERE { ?a  a  dbo:Athlete . ?a  dbp:goals  ?goals . OPTIONAL { ?a  foaf:homepage  ?page . } }
\end{lstlisting}
\end{example}
\vspace{-1em}
In order to initialize a local data store, i.e., the target dataset, SPARQL CONSTRUCT queries can be used by employing the interest expression's BGPs to extract and load a subset of the source dataset. 
Then interest expressions are registered with iRap to retrieve interesting updates from the source dataset. 
iRap evaluates interest expressions over changesets being published along with the source dataset. 
Without a restriction of generality, we assume interest expressions here to be static for the lifetime of a target dataset, since an evolution of interest expressions can be simulated by removal and addition.
The result of executing an interest evaluation for an interest expression against a changeset are three sets or triples: \textit{1. interesting, 2. potentially interesting, and 3. uninteresting} triples.

\begin{definition}[Interesting Triples] \label{def:InterestingTriples}
Interesting triples are all triples comprised in full matches of the BGP and possibly OGP of an interest expression, $i_g$, against the sets of added or deleted triples of a changeset. 
Interesting triples originating from the first element (i.e., removed triples ($D_{t_1-t_0}$)) of a changeset, $\Delta(V_{t_1})$, are called \textit{interesting-removed triples}. 
Interesting triples originating from the second element (i.e., added triples ($A_{t_1-t_0}$)) of a changeset, $\Delta(V_{t_1})$, are called \textit{interesting-added triples}.
\end{definition}

In addition to parts of an changeset for which the `interestingness' can be immediately decided, there might also be parts, which are \emph{potentially interesting} since, i) the missing parts to render them as interesting are already contained in the target knowledge base or ii) they will be propagated in subsequent updates.

\begin{definition}[Potentially Interesting Triples]\label{def:PITriples}
Potentially interesting triples are triples comprised in partial matches of the BGP or in OGP of interest expression, $i_g$:
\begin{itemize}[nosep]
	\item Potentially interesting triples originating from the first element (i.e., removed triples ($D_{t_1-t_0}$)) of a changeset $\Delta(V_{t_1})$, are called \textit{potentially interesting-removed triples}.
	\item Potentially interesting triples originating from the second element (i.e., added triples ($A_{t_1-t_0}$)) of a changeset, $\Delta(V_{t_1})$, are called \textit{potentially interesting-added triples}. 
\end{itemize}

\end{definition}
Potentially interesting triples can become interesting if triples missing in the changeset but required for a full BGP match are found in the target dataset or in subsequent changesets.
Finally, there are triples in the changeset that are neither interesting nor potentially interesting.

\begin{definition}[Uninteresting Triples]\label{def:UninterestingTriples}
Uninteresting triples are triples that do not match any triple pattern in a BGP or OGP of any interest expression, $i_g$, against the sets of added or deleted triples of a changeset. 
\end{definition}

Uninteresting triples are not interesting at the moment and can never become interesting with subsequent changesets.
iRap uses an interest query to select candidate triples from a changeset and to assert from a target dataset. 
These candidates are retrieved in decreasing order of matching BGP triple patterns of interest expressions and triples that match any part of optional graph patterns. 
Formal definition of \textit{interest candidate generation} from a changeset is:
\vspace{-0.5em}

\begin{definition}[Interest Candidate Generation] \label{def:IntCandidateGeneration}
An interest candidate generation is the extraction of matching triples from a changeset for a non-disjoint combination of triple patterns in BGP of an interest expression, $i_g$. 
The result of this extraction is an $(n+1)$-tuple with decreasing order of matching:
\vspace{-0.5em}
	\[ \pi(i_g, M) = \left\langle c_0, c_1, ..., c_{n-1}, c_{op}\right\rangle\] 
	\vspace{-0.2em}	
where:
\begin{itemize}[nosep]
	\item $M$ is a set of removed (respectively added) triples in a changeset,
	\item $n$ is the number of triple patterns in the BGP of interest expression, $i_g$, 
	\item $c_k$  is a set of candidate triples in $M$ that match $n-k$ $(0\leq k<n)$ triple patterns of the BGP (and optionally OGP) of the interest expression, $i_g$, and
	\item $c_{op}$ is a set of candidate triples in $M$ that match at least one triple pattern in the OGP of interest expression, $i_g$, but none of the triple patterns in the BGP.
\end{itemize}
\end{definition}
\begin{example}\label{exmp:candidateGeneration}
An interest candidate generation for the interest expression $i_g$ from~\autoref{exmp:interest} over the changeset from~\autoref{exmp:changeset} gives the following result:
\begin{enumerate}
\item
$\pi(i_g, D_{05:02-05:00}) = \left\langle c_0, c_1, c_{op}\right\rangle$ where:
\begin{description}[nosep]
\item[$c_0=\emptyset$]
\item[$c_1$] = \texttt{dbr:Marcel  dbp:goals  1.  dbr:Cristiano\_Ronaldo  dbo:goals  96.}
\item[$c_{op}=\emptyset$]
\end{description}
\item
$\pi(i_g, A_{05:02-05:00}) = \left\langle c_0, c_1, c_{op}\right\rangle$ where:
\begin{description}[nosep]
\item[$c_0$] = \texttt{dbr:Rio\_Ferdinand  a  dbo:Athlete .  dbr:Rio\_Ferdinand  dbp:goals  10. }
\item[$c_1$] = \texttt{dbr:Cristiano\_Ronaldo  dbp:goals  216 .  dbr:Arvid\_Smit  a  dbo:Athlete. }
\item[$c_{op}$] = \texttt{dbr:Barack\_Obama  foaf:homepage  "http://www.barackobama.com". }
\end{description}
\end{enumerate}
\end{example}

Now an interest candidate assertion verifies candidate triples with respect to all triple patterns in the BGP of an interest expression. 

\begin{definition}[Interest Candidate Assertion] \label{def:IntCandidateAssertion}
The candidate assertion function extracts missing triples for the candidate, $c_i$ of $\pi(i_g,M)$ of an interest expression $i_g$ from the target dataset, $\tau_{t_0}$:
\[ \pi'(i_g, M) = \left\langle c'_{op}, c'_{n-1}, ..., c'_1, c'_0\right\rangle \] where:
\begin{itemize}
\item $M$ is a set of removed (respectively added) triples in a changeset,
\item $n$ is the number of triple patterns in the BGP of interest expression, $i_g$,
\item $c'_{op}$ is a set of triples from target dataset, $\tau$, that matches the missing optional graph patterns for candidate $c_0$, of $\pi(i_g,M)$,
\item $c'_k$ is a set of triples from target dataset, $\tau$, that matches the missing triple patterns for candidate $c_{n-k}$, where $0< k < n$, of $\pi(i_g,M)$, and
\item $c'_0$ is a set of triples from target dataset, $\tau$, that matches all triple patterns in BGP of interest expression for candidate $c_{op}$, of $\pi(i_g,M)$.
\end{itemize}
\end{definition}
\begin{example}\label{exmp:candidateAssertion}
Let the target dataset, $\tau_{t_0}$,  at time $t_0$ contains the following triples:
\begin{lstlisting}
#Target dataset at time t0 = 05:00 PM Feb 06, 2015
dbr:Marcel              a              dbo:Athlete .
dbr:Marcel              dbp:goals      1 .
dbr:Cristiano_Ronaldo   a              dbo:Athlete .
dbr:Cristiano_Ronaldo   dbo:goals      96 .
dbr:Cristiano_Ronaldo   foaf:homepage  "http://cristianoronaldo.com" .
\end{lstlisting}
An interest candidate assertion for interest candidates generated  in~\autoref{exmp:candidateGeneration} yields the following result:
\begin{enumerate}
\item $\pi'(i_g, D_{05:02-05:00}) = \left\langle  c'_{op}, c'_1, c'_0\right\rangle$ where:
\begin{description}
\item[$c'_{op}=\emptyset$]
\item[$c'_1$] = \texttt{ dbr:Marcel              a              dbo:Athlete .\\
  dbr:Cristiano\_Ronaldo  a              dbo:Athlete .\\
  dbr:Cristiano\_Ronaldo  foaf:homepage  "http://cristianoronaldo.com" .}
\item[$c'_0=\emptyset$]
\end{description}
\item $\pi'(i_g, A_{05:02-05:00}) = \left\langle c'_{op}, c'_1, c'_0\right\rangle$ where:
\begin{description}
\item[$c'_{op}=\emptyset$] 
\item[$c'_1$] = \texttt{dbr:Cristiano\_Ronaldo  a              dbo:Athlete .\\
dbr:Cristiano\_Ronaldo  foaf:homepage  "http://cristianoronaldo.com" .}
\item[$c'_0=\emptyset$] 
\end{description}
\end{enumerate}
\end{example}
The interest evaluation over a changeset $\Delta(V_{t_1})$ is performed in two steps.
First, interest expressions are evaluated against removed triples of a changeset as $d(i_g, D_{t_1-t_0})$, see~\autoref{def:DeletionEvaluation}. 
Second, interest expressions are evaluated against added triples of a changeset as $\alpha(i_g, A_{t_1-t_0})$, see~\autoref{def:AdditionEvaluation}.
During interest evaluation, added triples are combined with potentially interesting triples from previous changesets (i.e., $I_{t_1-t_0} = A_{t_1-t_0} \cup \rho_{t_0}$) to check their potential promotion to interesting triples. 
\vspace{-0.5em}
\begin{definition}[Interest Evaluation over Deleted Triples] \label{def:DeletionEvaluation}
Interest evaluation over deleted triples is a function, $ d(i_g, D_{t_1-t_0})$, that returns a 3-element tuple\footnote{$\cup^*$ indicates that after the component-wise union of the two sets the results are combined to three categories of the resulting 3-tuple, namely, (i) elements from left that have matching right elements, (ii) elements from left that do not have matching right elements, and (iii) element from right that have a match left.}:
\vspace{-0.5em}
\[ d(i_g, D_{t_1-t_0}) = \pi(i_g, D_{t_1-t_0})\; \cup^* \; \pi'(i_g, D_{t_1-t_0}) =\left\langle r_{t_1-t_0}, r_{i(t_1-t_0)}, r'_{t_1-t_0}\right\rangle\] 
\vspace{-0.2em}
where:
\begin{itemize}[nosep]
\item $\pi(i_g, D_{t_1-t_0})$ is an interest candidate generation against deleted triples,
\item $\pi'(i_g, D_{t_1-t_0})$ is an interest candidate assertion against deleted triples,
\item $r_{t_1-t_0}=\{c_0 \cup c_k \cup c_{op} |\: c_0 , c_k, c_{op} \in \pi(i_g, D_{t_1-t_0}) \; and\; \exists c'_{n-k}, c'_0 \in \pi'(i_g, D_{t_1-t_0})\}$ is the set of interesting removed triples, i.e., no longer interesting,
\item $r_{i(t_1-t_0)}=\{c_k \cup c_{op} | c_k, c_{op} \in\pi(i_g, D_{t_1-t_0})\; and \: \nexists c'_{n-k}, c'_0\in \pi'(i_g, D_{t_1-t_0})\}$ is the set of \textit{potentially interesting removed triples} (existing only in removed triples of a changeset) and
\item $r'_{t_1-t_0} = \{c'_0 \cup c'_k \cup c'_{op}| c'_0, c'_k, c'_{op} \in  \pi'(i_g, D_{t_1-t_0}) \; and \; \exists c_{op}, c_{n-k}, c_0 \in  \pi(i_g, D_{t_1-t_0})$\} is the set of triples that become \textit{potentially interesting} after removing $r_{t_1-t_0}$. 
\end{itemize}
\end{definition}

\vspace{-0.5em}
\begin{example}\label{exmp:evalDeleted}
An interest evaluation over deleted triples in our running example (using the results of~\autoref{exmp:candidateGeneration} and \autoref{exmp:candidateAssertion}, respectively) is as follows: 
\vspace{-0.6em}
\begin{align*}
d(i_g, D_{05:02-05:00}) &= \pi(i_g, D_{05:02-05:00})\: \cup^* \: \pi'(i_g, D_{05:02-05:00}) \\
 &=\left\langle r_{05:02-05:00}, r_{i(05:02-05:00)}, r'_{05:02-05:00}\right\rangle 
\end{align*}
\begin{enumerate}
\item $r_{05:02-05:00}$ = $c_1$ (in ~\autoref{exmp:candidateGeneration})
\begin{lstlisting}[frame=none]
dbr:Marcel              dbp:goals    1 .
dbr:Cristiano_Ronaldo   dbo:goals    96 .                   
\end{lstlisting}
\vspace{-0.5em}
\item $r_{i(05:02-05:00)} = \emptyset$ (Since all the potentially interesting removed triples of $c_1$ in~\autoref{exmp:candidateGeneration} becomes interesting and no other triples in $c_op$)
\item $r'_{05:02-05:00}$ = $c'_1$
\begin{lstlisting}[frame=none]
dbr:Marcel              a              dbo:Athlete .
dbr:Cristiano_Ronaldo   a              dbo:Athlete .
dbr:Cristiano_Ronaldo   foaf:homepage  "http://cristianoronaldo.com" .
\end{lstlisting}
\end{enumerate}
\end{example}
\begin{definition}[Interest Evaluation over Added Triples]\label{def:AdditionEvaluation}
Interest evaluation over added triples is a function, $\alpha(i_g, A_{t_1-t_0})$, that returns 3 element tuple as: 
\vspace{-1em}
\[ \alpha(i_g, A_{t_1-t_0}) = \pi(i_g, I_{t_1-t_0})\; \cup^* \; \pi'(i_g, I_{t_1-t_0}) =\left\langle a_{t_1-t_0}, a_{i(t_1-t_0)}, a'_{t_1-t_0}\right\rangle \] 
where:
\begin{itemize}[nosep]
\item $I_{t_1-t_0} = A_{t_1-t_0} \cup \rho_{t_0}$ is a set of added triples and potentially interesting triples dataset,
\item $\pi(i_g, I_{t_1-t_0})$ is an interest candidate generation over $I_{t_1-t_0}$,
\item $\pi'(i_g, I_{t_1-t_0})$ is an interest candidate assertion over  $I_{t_1-t_0}$,
\item $a_{t_1-t_0}=\{c_0 \cup c_k \cup c_{op} |\: c_0 , c_k, c_{op} \in \pi(i_g, I_{t_1-t_0}) \; and\; \exists c'_{n-k}, c'_0 \in \pi'(i_g, I_{t_1-t_0})\}$ is the set of interesting added triples,
\item $a_{i(t_1-t_0)}=\{c_k \cup c_{op} | c_k, c_{op} \in\pi(i_g, I_{t_1-t_0})\; and \: \nexists c'_{n-k}, c'_0\in \pi'(i_g, I_{t_1-t_0})\}$ is the set of potentially interesting added triples that do not have related triples in target dataset, and
\item $a'_{t_1-t_0}=\{c'_0 \cup c'_k \cup c'_{op}| c'_0, c'_k, c'_{op} \in  \pi'(i_g, I_{t_1-t_0}) \; and \; \exists c_{op}, c_{n-k}, c_0 \in  \pi(i_g, I_{t_1-t_0}) \;$ respectively\}  is the set of triples from target dataset that are related to $a_{i(t_1-t_0)}$.
\end{itemize}
\end{definition}
\begin{example} \label{exmp:evalAdded}
An interest evaluation over added triples in our running example (using the results of~\autoref{exmp:candidateGeneration} and \autoref{exmp:candidateAssertion}, respectively) is as follows:
\vspace{-0.6em}
\begin{align*}
\alpha(i_g, A_{05:02-05:00})&=\pi(i_g, I_{05:02-05:00})\; \cup^* \; \pi'(i_g, I_{05:02-05:00})\\
&=\left\langle a_{05:02-05:00}, a_{i(05:02-05:00)}, a'_{05:02-05:00}\right\rangle
\end{align*}
\begin{enumerate}
\item $a_{05:02-05:00}$ = $c_1 \cup c'_1 \cup c_0$
\begin{lstlisting}[frame=none]
dbr:Cristiano_Ronaldo   dbo:goals      216 .
dbr:Cristiano_Ronaldo   a              dbo:Athlete .
dbr:Cristiano_Ronaldo   foaf:homepage  "http://cristianoronaldo.com" .     
dbr:Rio_Ferdinand       a              dbo:Athlete .
dbr:Rio_Ferdinand       dbp:goals      10 .                
\end{lstlisting}
\item $a_{i(05:02-05:00)}$ = 
\begin{lstlisting}[frame=none]
dbr:Arvid_Smit     a               dbo:Athlete .
dbr:Barack_Obama   foaf:homepage   "http://www.barackobama.com" .
\end{lstlisting}
\item $a'_{05:02-05:00}=\emptyset$
\end{enumerate}
\end{example}

Now, we will use the results from~\autoref{def:DeletionEvaluation} and~\autoref{def:AdditionEvaluation} to compute interesting and potentially interesting changesets.
\begin{definition}[Interest Evaluation] \label{def:InterestEvaluation}
An interest evaluation over a changeset $\Delta(V_{t_1})$ at time $t_1$ is a function $e(i_g, \Delta(V_{t_1}))$ that combines the results from an interest evaluation over deleted triples, $d(i_g, D_{t_1-t_0})$, and an interest evaluation over added triples, $\alpha(i_g, I_{t_1-t_0})$, to return an interesting changeset and potentially interesting changeset as follows:
\vspace{-0.5em}
\[e(i_g, \Delta(V_{t_1})) = d(i_g, D_{t_1-t_0}) \quad \chi \quad \alpha(i_g, I_{t_1-t_0}) = \left\langle \Delta(\tau_{t_1}), \Delta(\rho_{t_1})\right\rangle\] where
$i_g$ is an interest expression over an evolving dataset,
$\Delta(\tau_{t_1})$ is an interesting changeset (see~\autoref{def:InterestingChangeset}), and
$\Delta(\rho_{t_1})$ is potentially interesting changeset (see~\autoref{def:PInterestingChangeset}).
\end{definition}

\begin{definition}[Interesting Changeset]\label{def:InterestingChangeset}
Let $\tau_{t_0}$ be a target dataset at time $t_0$. An interesting changeset, $\Delta(\tau_{t_1})$, for $\tau_{t_0}$ at time $t_1$ is defined as:
\vspace{-0.5em}
\[ \Delta(\tau_{t_1}) = \left\langle \lgroup r_{t_1-t_0} \cup r'_{t_1-t_0}\rgroup, \; a_{t_1-t_0}\right\rangle\] where:
\begin{itemize}[nosep]
\item $r_{t_1-t_0}$ is the set of interesting removed triples, interesting removed optional triples and potentially interesting removed triples with match found in target dataset during candidate generation, $\pi(i_g, D_{t_1-t_0})$,
\item $r'_{t_1-t_0}$ is the set of triples from target dataset that are related to potentially interesting removed triples computed by $\pi'(i_g, D_{t_1-t_0})$, and
\item $a_{t_1-t_0}$ is the set of interesting added triples, interesting optional triples and potentially interesting added triples with match found in target dataset during candidate generation, $\pi(i_g, A_{t_1-t_0})$.
\end{itemize}
\end{definition}
\begin{example} \label{exmp:interestingChangeset}
An interesting changeset for our running example is as follows:
$\Delta(\tau_{05:02}) = \left\langle \lgroup r_{05:02-05:00} \cup r'_{05:02-05:00}\rgroup, \; a_{05:02-05:00}\right\rangle$
\begin{enumerate}
\item interesting removed triples -- $\lgroup r_{05:02-05:00} \cup r'_{05:02-05:00}\rgroup$ :
\begin{lstlisting}
dbr:Marcel             a              dbo:Athlete .
dbr:Marcel             dbp:goals      1 .
dbr:Cristiano_Ronaldo  dbo:goals      96 .
dbr:Cristiano_Ronaldo  a              dbo:Athlete .
dbr:Cristiano_Ronaldo  foaf:homepage  "http://cristianoronaldo.com" .                     
\end{lstlisting}
\item interesting added triples --  $a_{05:02-05:00}$ :
\begin{lstlisting}
dbr:Cristiano_Ronaldo   dbo:goals      216 .
dbr:Cristiano_Ronaldo   a              dbo:Athlete .
dbr:Cristiano_Ronaldo   foaf:homepage  "http://cristianoronaldo.com" .     
dbr:Rio_Ferdinand       a              dbo:Athlete .
dbr:Rio_Ferdinand       dbp:goals      10 .  
\end{lstlisting}  
\end{enumerate}
\end{example}
\begin{definition}[Potentially Interesting Changeset]\label{def:PInterestingChangeset}
Let $\rho_{t_0}$ be a potentially interesting dataset for interest expression $i_g$ at time $t_0$. A changeset, $\Delta(\rho_{t_1})$, for $\rho_{t_0}$ at time $t_1$ is defined as:
\vspace{-0.5em}
\[ \Delta(\rho_{t_1}) = \left\langle r_{i(t_1-t_0)}, \; \lgroup a_{i(t_1-t_0)} \cup r'_{t_1-t_0}\rgroup\right\rangle \] where:
\begin{itemize}[nosep]
\item $r_{i(t_1-t_0)}$ is a set of potentially interesting removed triples,
\item $a_{i(t_1-t_0)}$ is a set of potentially interesting added triples computed on added triples of a changeset and related triples extracted from target while removing potentially interesting removed triples, and
\item $r'_{t_1-t_0}$ is the set of triples from target dataset that are related to potentially interesting removed triples computed by $\pi'(i_g, D_{t_1-t_0})$.
\end{itemize}
\end{definition}
\begin{example} \label{exmp:PIChangeset}
Potentially interesting changeset for our running example is as follows: $\Delta(\rho_{05:02}) = \left\langle r_{i(05:02-05:00)}, \; \lgroup a_{i(05:02-05:00)} \cup r'_{05:02-05:00}\rgroup\right\rangle  $
\begin{enumerate}
\item Potentially interesting removed triples -- $r_{i(05:02-05:00)} = \emptyset$
\item Potentially interesting added triples -- $\lgroup a_{i(05:02-05:00)} \cup r'_{05:02-05:00}\rgroup$
\begin{lstlisting}[frame=single]
dbr:Arvid_Smit    a              dbo:Athlete .
dbr:Barack_Obama  foaf:homepage  "http://www.barackobama.com" .
dbr:Marcel        a              dbo:Athlete .
\end{lstlisting}
\end{enumerate}
\vspace{-1em}
\textbf{Note:} since all triples in $r'_{05:02-05:00}$ are added back to target dataset, they are no longer stored in the potentially interesting dataset.
\end{example}
\begin{definition}[Interesting Update Propagation]\label{def:InterestPropagation}
An interesting changeset propagation is an update operation  that transforms  the target dataset $\tau_{t_0}$ to the new dataset $\tau_{t_1}$ and $\rho_{t_0}$ to new dataset $\rho_{t_1}$ by applying the result of interest evaluation, $e(i_g, \Delta(V_{t_1}))$. That is:
\[ \Upsilon(i_g, \Delta(V_{t_1})) =   \upsilon(\tau_{t_0},\Delta(\tau_{t_1})) \; \land\; \upsilon(\rho_{t_0},\Delta(\rho_{t_1}))  =  \tau_{t_1}\; \land \;  \rho_{t_1}  \]
\begin{itemize}[nosep]
\item $\Delta(V_{t_1})$ is a changeset at time $t_1$,
\item $\upsilon(\tau_{t_0},\Delta(\tau_{t_1}))=\lgroup \tau_{t_0} \backslash [ r_{t_1-t_0} \cup r'_{t_1-t_0}]\rgroup \cup a_{t_1-t_0} $ is changeset propagation of interesting changeset, and
\item $\upsilon(\rho_{t_0},\Delta(\rho_{t_1}))= \lgroup\rho_{t_0}\backslash r_{i(t_1-t_0)}\rgroup \cup \lgroup a_{i(t_1-t_0)} \cup r'_{t_1-t_0}\rgroup$ is changeset propagation of potentially interesting changeset.
\end{itemize}
\end{definition}

\begin{example} \label{exmp:propagation}
Propagation of an interesting changeset of~\autoref{exmp:interestingChangeset} to the target dataset, $\tau_{t_0}$ and potentially interesting changeset of~\autoref{exmp:PIChangeset} to the potentially interesting dataset$\rho_{t_0}$ transforms the datasets to:
\vspace{-0.5em}
\begin{figure}[tbh]
\begin{subfigure}[h]{0.498\textwidth}
\begin{lstlisting}[caption={Resulting target dataset}]
dbr:Cristiano_Ronaldo  dbo:goals      216 .
dbr:Cristiano_Ronaldo  a              dbo:Athlete .
dbr:Cristiano_Ronaldo  foaf:homepage  
                    "http://cristianoronaldo.com" . 
dbr:Rio_Ferdinand      a              dbo:Athlete . 
dbr:Rio_Ferdinand      dbp:goals      10 .  
\end{lstlisting}
\end{subfigure}
\hspace{.2cm}
\begin{subfigure}[h]{0.48\textwidth}
\begin{lstlisting}[caption={Potentially interesting dataset after change propagation}]
dbr:Arvid_Smit    a              dbo:Athlete .
dbr:Barack_Obama  foaf:homepage  
                 "http://www.barackobama.com" .
dbr:Marcel        a              dbo:Athlete .
\end{lstlisting}
\end{subfigure}
\end{figure}
\end{example}

\vspace{-1em}
\section{iRap RDF Update Propagation Framework} \label{sec:iRapImpl}
\vspace{-1em}
In this section we describe the architecture of our interest-based update propagation framework iRap and its implementation. 
iRap was implemented in Java using Jena-ARQ.  
It is available as open-source\footnote{\url{http://eis.iai.uni-bonn.de/Projects/iRap}} and consists of three modules: 
(1) \textit{Interest Manager} (IM), 
(2) \textit{Changeset Manager} (CM) and 
(3) \textit{Interest Evaluator} (IE), each of which each can be extended to accommodate new or improved functionality.

\begin{figure}[h]
	\centering	
	\includegraphics[width=.8\textwidth]{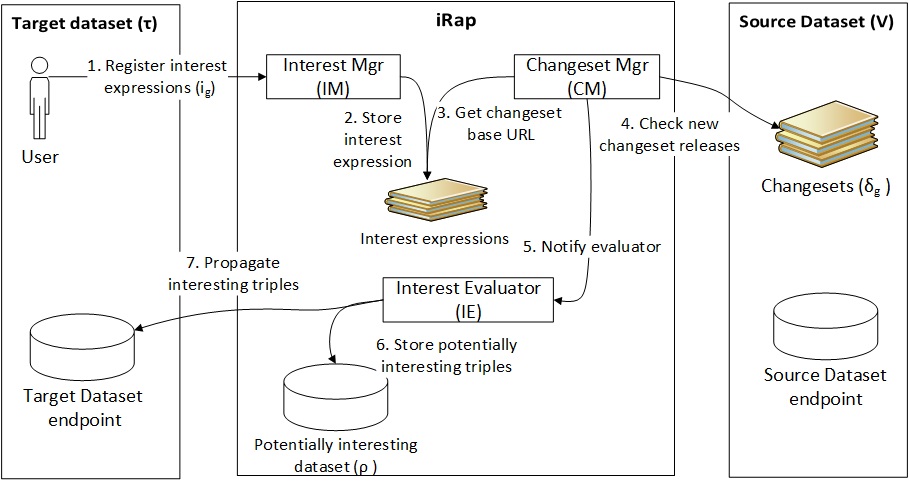}
	\caption{Architecture of the iRap interest-based RDF update propagation framework.}
	\label{fig:arch}
\end{figure}

Changeset evaluation starts after a user registers an interest expression using the IM service, as shown in ~\autoref{fig:arch}. 
The CM module fetches a list of changeset folders from interest expressions and regularly (configurable) checks for new changesets. 
After downloading and decompressing new changesets, the CM notifies the IE, which then imports a list of interest expressions registered for this particular changeset through the IM and initiates the evaluation.
Resulting interesting triples are propagated to the target dataset whereas potentially interesting triples are stored in the potentially interesting dataset ($\rho$). 
After all interest expressions have been evaluated over the changeset, the IE notifies the CM to clean the downloaded files.

\vspace{-1em}
\section{Evaluation}\label{sec:evaluation}
\vspace{-1em}
To evaluate the proposed approach, we performed experiments on the iRap framework using changesets published by DBpedia and compared the results with the DBpedia Live Mirror tool. 
The comparison considers two cases: using iRap to update a previously-established local replica of i) an entire remote dataset ii) a subset of a remote dataset. These two cases simulate two ways in which iRap can be used: i) using interest-based changeset propagation for future updates of a local copy of a large dataset or ii) starting with a new subset of the large dataset.

\subsubsection{Experimental Setting}
In order to test our approach we used the DBpedia dump\footnote{\url{http://live.dbpedia.org/dumps/dbpedia_2014_09_30_00_00.fixed.ttl.gz}} of September 30, 2014 for the initial setup of the target datasets for two different application domains, namely, \textit{Location and Football} datasets.
Changesets published between October 01 and October 15, 2014 (see \autoref{tbl:dbchanges}) were used for evaluation\footnote{\url{http://live.dbpedia.org/changesets/2014/10/}}.
Initially we set up two TDB datasets for each target dataset from the DBpedia dump. 
We loaded all triples from the dump to the Location dataset, whereas for the Football dataset we only loaded slice corresponding to interesting triples matching \autoref{qry:footballQuery}.

Initially, the Location dataset contains all triples from DBpedia yielding a total of 364,810,370 triples, whereas the Football dataset contains only 265,622 triples.
A total of 12,057 changesets (pairs of removed and added .nt.gz files) have been published in the evaluation timeframe.
\begin{table}[tb]
 \vspace{-1.5em}
\centering 
\scriptsize
    \begin{tabular}{| c | c | c | c | c    | c | c | c |}
    \hline
    \textbf{Date} & \textbf{Oct 01} & \textbf{Oct 02} & \textbf{Oct 03}  & \textbf{Oct 04-12}& \textbf{Oct 13} & \textbf{Oct 14} & \textbf{Oct 15}   \\ \hline
    \textbf{Total Changesets}	& 0 & 1,621 & 1,755  &  0 & 5,352 & 751 & 2,578  \\   
    \hline	
    \end{tabular}
    \caption{Distribution of DBpedia Live changesets published October 01-15, 2014.}
    \label{tbl:dbchanges}    
 \vspace{-3em}
 \end{table}
 
The evaluation comprises two interest expressions, $I_1$ and $I_2$. 
$I_1$ comprises a non-disjoint BGP containing 4 triple patterns with a maximum of two variables per triple pattern (object-subject join)~\autoref{qry:footballQuery}.
$I_2$ comprises a non-disjoint BGP containing 5 triple patterns with a maximum of two variables per triple pattern (subject-subject joins) and one an OGP containing one triple pattern~\autoref{qry:locationQuery}.

\begin{figure}[tb]
\begin{subfigure}[h]{0.48\textwidth}
\begin{lstlisting}[language=sparql,escapechar=@,morekeywords={STRSTARTS}, caption={Location interest query}, label={qry:locationQuery}]
CONSTRUCT WHERE {
  ?location  a             ?type .
  ?location  wgs:long      ?long .
  ?location  wgs:lat       ?lat .
  ?location  rdfs:label    ?label .
  ?location  dbo:abstract  ?abstract .
  OPTIONAL { ?location dcterms:subject ?subject }
}
\end{lstlisting}
\end{subfigure}
\hspace{.2cm}
\begin{subfigure}[h]{0.48\textwidth}
\begin{lstlisting}[language=sparql,escapechar=@,morekeywords={STRSTARTS}, caption={Football interest query}, label={qry:footballQuery}]
CONSTRUCT WHERE {
  ?footballer  a    dbo:SoccerPlayer .
  ?footballer  foaf:name   ?name.
  ?footballer  dbo:team    ?team .
  ?team        rdfs:label  ?teamName. 
}
\end{lstlisting}
\end{subfigure}
\end{figure}
We set up two target datasets and potentially interesting dataset using Jena TDB and jena-fuseki for each dataset.
The potentially interesting dataset stores potentially interesting triples for each interest expression within a named graph. 
All experiments were carried out on a 64-bit machine with Windows 7, Intel(R) Core i7-4770 CPU, 16GB RAM and 1TB HD.

\subsubsection{Evaluation Results and Discussion}
\begin{figure}[tb]
\vspace{-1.5em}
        \centering
        \begin{subfigure}[tb]{0.49\textwidth}
                \includegraphics[width=\textwidth]{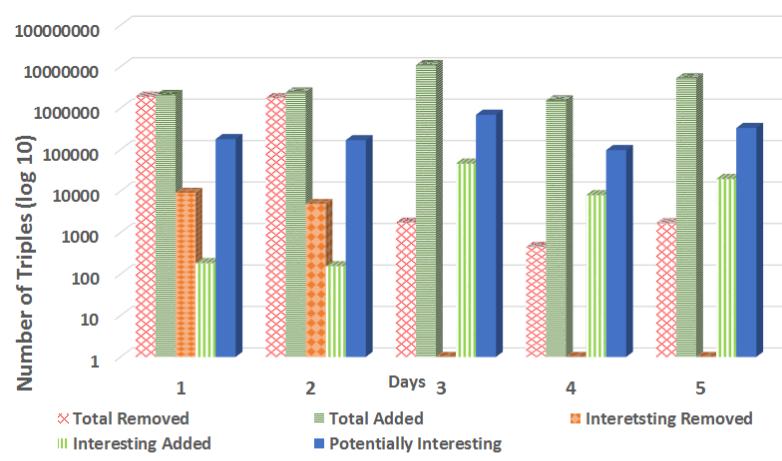}
                \caption{Football dataset changes per day}
                \label{fig:footballEvalPerDay}
        \end{subfigure}
        \begin{subfigure}[tb]{0.49\textwidth}
                \includegraphics[width=\textwidth]{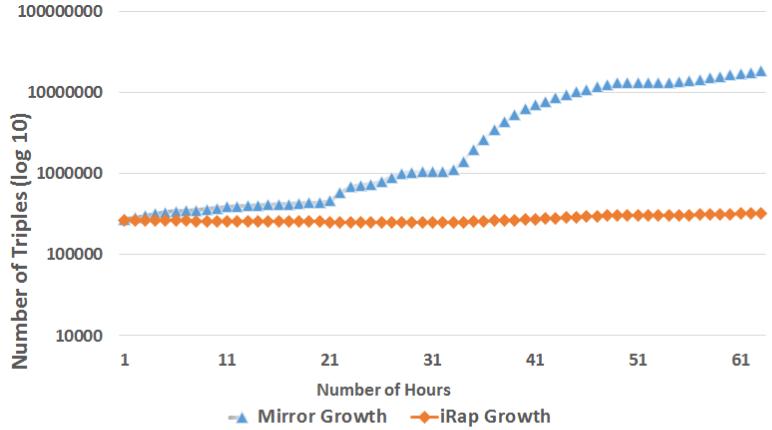}
               \caption{Football dataset growth}
               \label{fig:footballDsGrowth}
       \end{subfigure}
        \begin{subfigure}[tb]{0.49\textwidth}
                \includegraphics[width=\textwidth]{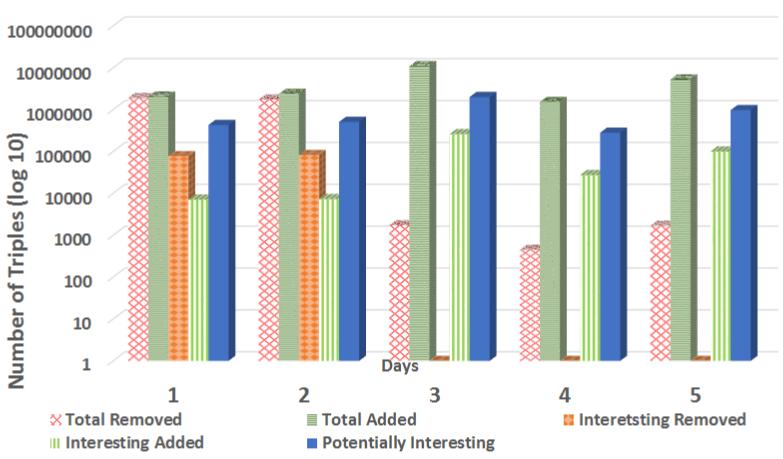}
                \caption{Location dataset changes per day}
                \label{fig:locationEvalPerDay}
        \end{subfigure}
        \begin{subfigure}[tb]{0.49\textwidth}
                \includegraphics[width=\textwidth]{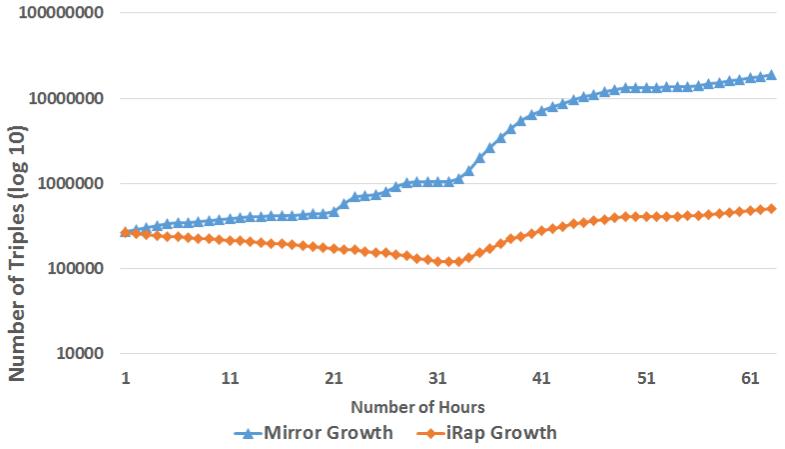}
               \caption{Location dataset growth}
               \label{fig:locationDsGrowth}
       \end{subfigure}
        \begin{subfigure}[tb]{1\textwidth}
                \includegraphics[width=\textwidth]{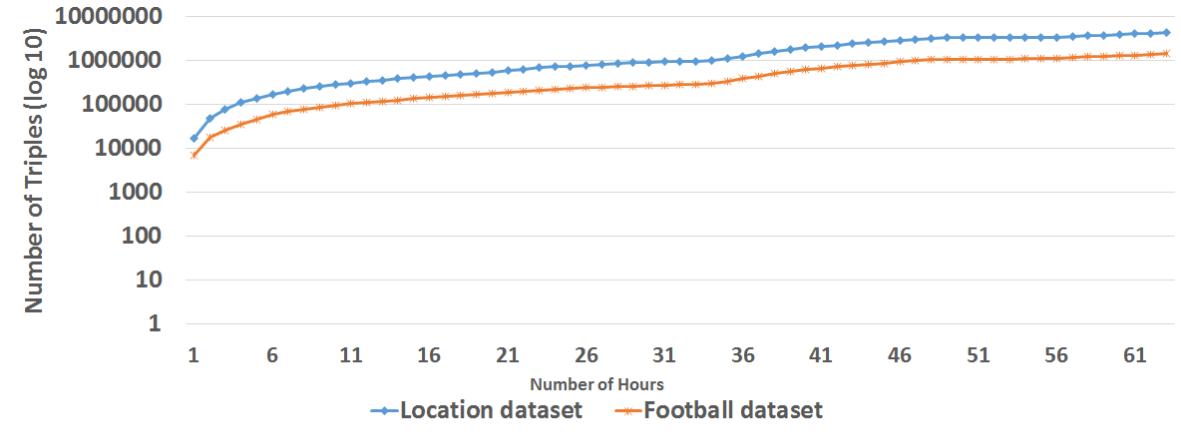}
               \caption{Potentially interesting dataset growth}
               \label{fig:piGrowthForBoth}
               \vspace{-1em}
       \end{subfigure}     
        \caption{Evaluation results}
        \label{fig:results}
\vspace{-1.5em}
\end{figure}

\autoref{fig:results} summarizes our experimental results for two target datasets shows the growth of the potentially interesting dataset. 
Results of the interest evaluation for the Football dataset are presented in~\autoref{tbl:soccer}. 
From the overall changesets considered for this evaluation, in~\autoref{tbl:dbchanges}, only 0.38\% of the removed and 0.335\% of the added triples were identified as interesting for the Football dataset.
The average changeset publication interval was 18.81s and average time required for a changeset evaluation is 0.87s.
This shows that iRap efficiently performs changeset propagations way before the next changeset is published.

\begin{table}[tb]
\vspace{-1em}
\centering
\scriptsize
    \begin{tabular}{| c | c | c | c | c | c | c |}
    \hline
    \textbf{Day} & \textbf{Total} & \textbf{Interesting} & \textbf{Total}  & \textbf{Interesting} & \textbf{Potentially} & \textbf{Elapsed} \\ 
		    & \textbf{Removed} & \textbf{Removed} & \textbf{Added} & \textbf{Added} & \textbf{Interesting} & \textbf{(in minutes)} \\ \hline
    1	& 1,895,179  & 9,065   & 2,051,976	 & 184	   & 169,554  & 15.18 \\ \hline
	2	& 1,748,511	 & 4,865   & 2,384,232	 & 155	   & 168,856  & 20.85 \\ \hline
	3	& 1,716	     & 0	   & 10,728,855  & 45,429  & 684,491  &  69.86 \\ \hline
	4	& 449        & 0	   & 1,522,939	 & 7,970   & 97,300   & 10.17 \\ \hline
	5	& 1,677	     & 0	   & 5,234,788	 & 19,598  & 333,232  & 60.06\\ \hline
    \end{tabular}
    \caption{Comparison of results for Football App}
    \label{tbl:soccer}
 \end{table}

Results of the interest evaluation for the Location dataset are shown in~\autoref{tbl:location}. 
From the overall changesets considered for this evaluation, in~\autoref{tbl:dbchanges}, only 4.38\% of the removed and 1.81\% of the added triples were interesting for the Location dataset.
The average time spent for a changeset evaluation is 5.31s.
The interest evaluation for the Location dataset takes longer than Football dataset, because of the number of triples in the target dataset was a the full DBpedia.

\begin{table}[tb]
\vspace{-1em}
\centering
\scriptsize
    \begin{tabular}{| c | c | c | c | c | c | c |}
    \hline
    \textbf{Day} & \textbf{Total} & \textbf{Interesting} & \textbf{Total}  & \textbf{Interesting} & \textbf{Potentially} & \textbf{Elapsed} \\ 
		    & \textbf{Removed} & \textbf{Removed} & \textbf{Added} & \textbf{Added} & \textbf{Interesting} & \textbf{(in minutes)} \\ \hline
    1	& 1,895,179  & 77,377	& 2,051,976	 & 7,093    & 430376     & 166.59 \\ \hline
	2	& 1,748,511	 & 82,461   & 2,384,232	 & 7,301    & 509,972    & 242.62 \\ \hline
	3	& 1,716	     & 0	    & 10,728,855 & 259,587  & 2,002,271  & 417.87 \\ \hline
	4	& 449        & 0	    & 1,522,939	 & 27,292   & 280,718    & 64.41 \\ \hline
	5	& 1,677	     & 0	    & 5,234,788	 & 100,073  & 972,284    & 176.78 \\ \hline
    \end{tabular}
    \caption{Comparison of results for Location App}
    \label{tbl:location}
\vspace{-2em}
 \end{table}

\autoref{fig:footballEvalPerDay} shows the number of triples published per a day and the number of interesting triples and potentially interesting triples found from interest evaluation for Football dataset. 
\autoref{fig:footballDsGrowth} shows the dataset growth comparison between iRap and a full mirror approach. 
As the figure clearly shows, iRap managed datasets are almost two orders of magnitude smaller and grow much slower than with a mirror approach. 
Note that the growth for each datasets is calculated by subtracting the number of removed triples from and adding the number of added triples to the total number of triples in the dataset.

\autoref{fig:piGrowthForBoth} shows a substantial growth of potentially interesting dataset for Location  and Football datasets. 
This is due to the number of variables used in triple patterns, and the number and type of triple patterns in interest expression. 
For example, the Football dataset interest query contains the common predicates \texttt{foaf:name} and \texttt{rdfs:label} which are used in almost all resources and thus result in many potentially interesting triples. 
Exploring further options to reduce the growth of the potentially interesting dataset is thus an interesting direction for future work.
Again, the average processing time per changeset is always way below the average time between two changesets.
The correctness of the resulting triples from the first changesets, for Football dataset interest expression, was checked by manual inspection.

\vspace{-1em}
\section{Related Work}\label{sec:relatedwork}
\vspace{-1em}
Most related work on dataset change detection and propagation focuses on distributed publish/subscribe systems~\cite{Pellegrino2013,Chirita2004}, resource link maintenance~\cite{PopitschH11,tramp2010weaving}, target synchronization~\cite{Tummarello2007}, partial replicas~\cite{Schandl2010}, data-shipping~\cite{Voruganti_anadaptive}, lazy updates~\cite{Breitbart1999}, and real-time update notification~\cite{tramp2010weaving,passant2010sparqlpush}. 
In~\cite{Pellegrino2013}, the authors propose a peer-to-peer publish/subscribe system for events described in RDF.
By avoiding the use of multiple indexes for the same publication they manage to reduce storage space.
Similarly, \cite{Chirita2004} provide an implementation with publish/subscribe capabilities in an RDF-based peer-to-peer system to manage digital resources. 
As for resource link maintenance, \emph{DSNotify}~\cite{PopitschH11} offers a change-detection framework to detect and fix broken links between resources in two datasets while, \emph{Semantic Pingback}~\cite{tramp2010weaving} proposes a notification system for the creation of new links between Web resources.
To note that this approach is suitable for relatively static resources, i.e. RDF documents or RDFa annotated Web pages.
In contrast, \emph{SparqlPuSH}~\cite{passant2010sparqlpush} offers a real-time notification framework for data updates in a RDF store using a semantic \emph{PubSubHubbub}-based protocol (PuSH). 
SparqlPuSH allows users to subscribe for changes updates of a subset of content in a RDF store using SPARQL.
However, notification and broadcasting are only available as RSS and Atom feeds.
As regards target synchronization, \emph{RDFSync}~\cite{Tummarello2007}
performs update synchronization by merging source and target graphs to get the updated target RDF graph.
Alternatively, \cite{Schandl2010} has designed an approach to replicate, modify, and write-back parts of an RDF graph on devices with low computing power. 
However, this approach does not resolve conflicts arising with concurrent modifications on both the base graph and the partial replicas.
In the field of object database management systems, a data-shipping client-server architecture, such as in ~\cite{Voruganti_anadaptive}, is used for data distribution. The aim is to optimize resource utilization at client side where the data objects from the server are cached for future use. 
In distributed databases, where data is replicated on different sites, Lazy update protocols~\cite{Breitbart1999} disseminate updates to replicas to ensure consistency. 
These protocols guarantee serializable execution as well as high performance.

\section{Conclusion and Future Work}\label{sec:conclusion}
\vspace{-1em}
In this paper we presented a novel approach for interest-based RDF update propagation that can consistently maintain a full or partial replication of large LOD datasets. 
We have demonstrated the validity of the approach through detailed formalizations and their application in a reference implementation of the iRap Framework. 
An thorough evaluation of the approach, using large-scale real-world data dumps and changesets regularly provided by a renowned LOD dataset, indicates that our method can significantly cut down on both the size of the data updates required to consistently maintain a localized dataset replication up-to-date, as well as the speed by which such updates can take place. 

Future work will focus on extending the iRap Framework with a publish/subscribe distributed architecture as described in the related work (Section~\ref{sec:relatedwork}).
The framework will be improved also from the usability point of view, including a user interface and making the initial generation of RDF slices easier and more efficient. 
Finally, an extensive evaluation of scalability and performance of the framework will be performed and a benchmark dataset for future reference will be made available to the research community.

\vspace{-1em}
\bibliographystyle{plain}
\bibliography{irap}

\end{document}